# When linear inversion fails: neural-network optimization for sparse-ray travel-time tomography of a volcanic edifice

**Abolfazl Komeazi[1*], Georg Rümpker[1,2], Johannes Faber[2], Fabian Limberger[1], Nishtha Srivastava[1,2]**

[1]Institute of Geosciences, Goethe University Frankfurt, Frankfurt, Germany
[2]Frankfurt Institute for Advanced Studies, Frankfurt, Germany

**\* Correspondence:**
Abolfazl Komeazi
komeazi@geophysik.uni-frankfurt.de

**Keywords: Volcanic edifice, Neural Network, Machine Learning, Magma chamber, Tomography, Inversion**

## Abstract

In this study, we present an artificial neural network (ANN)-based approach for travel-time tomography of a volcanic edifice. We employ ray tracing to simulate the propagation of seismic waves through the heterogeneous medium of a volcanic edifice, and an inverse modeling algorithm that uses an ANN to estimate the velocity structure from the "observed" travel-time data. The performance of the approach is evaluated through a 2-dimensional numerical study that simulates i) an active source seismic experiment with few (explosive) sources placed on one side of the edifice and a dense line of receivers placed on the other side, and ii) earthquakes located inside the edifice and receivers placed on both sides of the edifice. The results are compared with those obtained from conventional damped linear inversion, demonstrating that the ANN-based approach outperforms the classical approach, particularly in situations with sparse ray coverage. Our study emphasizes the advantages of employing a relatively simple ANN architecture in conjunction with second-order optimizers to minimize the loss function. The ANN-based approach is computationally efficient and capable of providing high-resolution velocity images of anomalous structures within the edifice, making it a potentially valuable tool for the detection of low velocity anomalies related to magmatic intrusions or mush.

## 1 Introduction

Travel-time tomography has been shown to be a valuable tool for gaining insights into the spatial distribution of volcanic conduits, magma chambers, and other volcanic subsurface structures that are crucial for enhancing our understanding of magmatic processes (e.g., Lees, 2007; Magee et al., 2018; D'Auria et al., 2022). Studies aiming to image the subsurface at active volcanoes (e.g., Benz et al., 1996; Okubo et al., 1997; Lees and Crosson, 1989, 1990; Nakamichi, 2007; Paulatto et al., 2010; Shalev et al., 2010) have reported velocity anomalies in the shallow and deeper crust, which have been partially explained by remnants of historic magmatism or solidified andesitic cores of the volcano complexes, as well as by melt accumulations within dykes, sills or magma chambers, respectively. However, accurately imaging the interior of a volcanic edifice can be challenging due to the complex and highly heterogeneous velocity structure of the subsurface and sparse data sets, which results in considerable ambiguity in the interpretations (e.g., see review by Lees, 2007; Zandomeneghi et al., 2013).

Previous studies (e.g., see Magee et al., 2018 for a general review; Hammond and Kendall, 2016) found that the observed waveforms from raypaths through the edifice are mostly influenced by an accumulative effect of magma-filled intrusions or hydrothermal dykes or sills. However, the discrimination between these effects by traditional seismic tomography methods remains difficult. Furthermore, the limited ray coverage of seismic networks in volcanic regions can lead to inaccuracies in the inversion results (e.g., Stork and Clayton, 1991; Neumaier, 1998; Tarantola, 2005). This problem emerges when the number of unknowns (model parameters describing the velocity structure), is considerably larger than the number of rays (which determines the number of relevant equations). The resulting ill-posed nature of the inversion problem can yield multiple potential solutions and thus adds complexity to the interpretation and understanding of the seismic data.

To address these challenges, several studies have proposed the use of non-linear inversion techniques, such as regularization and Bayesian inversion, to improve the accuracy and robustness of travel-time tomography (e.g., Tarantola and Valette, 1982; Mosegaard and Tarantola, 1995; Virieux and Operto, 2009; Zhu et al., 2016; Fichtner and Zunino, 2019; Zhao et al., 2021). While these approaches serve to mitigate certain drawbacks of linear inversion methods, limited ray coverage may still lead to significant smearing of velocity anomalies along raypaths. Additionally, a reliable initial velocity model is crucial for the success of these methods, as a poor approximation of the initial velocity model may prevent convergence of the inversion. Recently, Artificial Neural Networks (ANNs) have shown promise in addressing these challenges of seismic tomography. ANN-based approaches have been used to improve the resolution and accuracy of velocity models in various geological settings, some of them including volcanic regions (e.g., Araya-Polo et al., 2018; Wang et al., 2020; O'Brien et al., 2023; Taufik et al., 2023). These approaches utilize the non-linear mapping capabilities of ANNs to capture complex relationships between the travel-time (and waveform) data and the subsurface velocity structure, which can be difficult to model by using traditional linear inversion techniques. Nevertheless, prior studies have not assessed the performance of neural networks in scenarios characterized by sparse ray coverage when linear inversion fails to provide a meaningful image of the subsurface.

In this study, our primary focus is on the development and optimization of an ANN designed to detect velocity anomalies within a volcanic edifice by mapping the travel times of rays passing through a model represented by its P-wave velocity structure. We place particular emphasis on scenarios with limited ray coverage. Furthermore, we evaluate the performance of both first and second-order optimizers to constrain the parameters of the ANN. The latter has the potential to substantially improve the robustness and stability of the inversion results.

## 2 Model and methods

To develop and to test the performance of our approach, we generated synthetic data using a straightforward and realistic configuration. In order to minimize the impact of parametrization on the results and maintain the focus on comparing the performance between traditional linear inversion and ANN, we consider several simplifying assumptions: 1) Raypaths are calculated for a homogeneous background velocity structure. Therefore, we do not account for reflection, refraction, and diffraction phenomena. The straight-ray assumption allows the application of a one-step linear inversion approach that we use for comparison. 2) The location of the sources is assumed to be known a priori. a. 3) Effects of noise are not considered to only focus on differences that result from the different methodologies. 4) To better focus on the advantages of the ANN approach over damped linear inversion (in cases of sparse ray coverage) and to minimize computational costs, we implement our simulations in 2 dimensions.

## 2.1 Edifice model setup and forward simulation

To model the edifice, we use a gridded rectangle with dimensions of 2.7 km in length and 1.15 km in height (see Figure S1). The grid points are spaced at intervals of 0.05 km in both the x and z directions. The shape of the edifice is mimicked by a Gaussian function. A homogeneous background velocity of 1.5 km/s is considered, and low velocity anomalies of ~30% are created on top of the background velocity to serve as a model for zone of magma accumulations. The average velocity of 1.5 km/s is based on the P-wave velocity estimates obtained at the edifice of Oldoinyo Lengai volcano (Rümpker et al., 2022). We use the following formulation of the forward problem to generate the travel times

$$d = G.m \quad (1)$$

where $d$ represents the travel-times vector (data) for all rays, $G$ is a matrix of ray segments for each block of the medium, and $m$ is the vector of slowness values (1/velocity) for each block. In matrix form equation (1) can be expressed as

$$\begin{bmatrix} d_1 \\ d_2 \\ \vdots \\ d_K \end{bmatrix} = \begin{bmatrix} G_{11} & G_{12} & \cdots & G_{1L} \\ G_{21} & G_{22} & \cdots & G_{2L} \\ \vdots & \vdots & \vdots & \vdots \\ G_{K1} & G_{K2} & \cdots & G_{KL} \end{bmatrix} . \begin{bmatrix} m_1 \\ m_2 \\ \vdots \\ m_L \end{bmatrix} \quad (2)$$

where K represents the number of rays (equal to the number of sources multiplied by the number of receivers, Ns*Nr), and L represents the number of model parameters (equal to the number of blocks, Nb). Elementwise multiplication of $G$ by $m$ yields the travel time for the corresponding block. Subsequently, each element in $d$ represents the summation of travel times for all blocks along the raypath between a source-receiver pair.

## 2.2 Travel-time tomography using linear inversion

In the inverse problem concerning equation (1), the objective is to determine the model parameters based on the given travel times. This requires computation of the inverse of $G$ which is generally not quadratic. To address this, we calculate the generalized inverse of G by multiplying both sides of equation 1 by the transpose of G (Nolet, 2008), which yields

$$m = (G^T G)^{-1}. G^T. d \quad (3)$$

In seismic tomography, when ray coverage is limited, $G^T G$ is a sparse matrix and due to vanishing of the determinant, computation of the inverse matrix becomes unstable. We, therefore, apply a standard damped least-squares method (Lines and Treitel, 1984), which leads to

$$m = (G^T G + \lambda^2 I)^{-1}. G^T . d \quad (4)$$

where λ is the damping factor, and *I* is the unit matrix. In this study, a value of 0.001 for λ was determined through experimentation.

## 2.3 Machine learning approach

We employ a supervised machine learning technique using a feedforward ANN (hereafter ANN) to act as the inverse operator. ANNs consist of several layers of weighted neurons, as shown in Figure 1. The initial layer is the input layer, followed by one or more hidden layers which are finally connected to the output layer. In this approach, ANN maps vectors of travel times to corresponding velocity model vectors, with travel times serving as inputs and velocity models as labels. This enables the ANN to "predict" the velocity model for a set of travel times based on an unseen velocity distribution.

The training process can be viewed as repetitive iterations aimed at minimizing a loss function. Selection of appropriate hyperparameters is crucial in this approach and is typically achieved through grid search over the hyperparameter space. In this study, we use the Mean Squared Error (MSE) as the loss function, which is a common choice for regression problems. The tanh activation function is applied to the hidden layers (as it can more effectively manage the curvature information conveyed by second-order derivative optimizers), and L-BFGS is chosen as optimizer (this choice will be discussed further in section 3). We implemented the machine learning approach using PyTorch version 2.0.1 (Paszke et al., 2019) on a single NVIDIA GeForce RTX 2080 Ti GPU.

To train the ANN, we create 10000 random velocity models and use equation (1) to calculate the corresponding travel times for each model. 70% of these models are used as training dataset, 15% are used as validation dataset and the remaining 15% serve as test dataset. The training process takes about 10 minutes to be accomplished on a single GPU. Examples of typical training velocity models are depicted in Figure 2. For all models, the velocities in each block are chosen randomly between 1.0 and 2.0 km/s. This choice of velocities minimizes the correlation not only between the training datasets and the validation and test models but also between the training datasets and the example models (described in the following) to assess the performance of the ANN. In theory, given sufficient training models, ANN will be able to simulate the inverse of *G* using statistical learning (Hastie et al., 2009).

After an initial evaluation of a variety of ANN architectures by trial and error, we realized that a relatively simple model, consisting of two hidden layers (Figure 1), in combination with a second-order derivative ANN optimizer resulted in a better performance regarding the minimization of the loss function and calculation time. Figures S2 and S3 show examples for the performance of the different architectures by visualizing the training evolution through epochs. These graphs clearly indicate that employing additional neurons or adding more hidden and convolutional layers leads to delayed learning by the ANN, without significant improvement in decreasing the loss function.

In this study, we focus on evaluating the performance of a deep machine learning approach using synthetic case studies, and we compare its results to those obtained from conventional linear inversion. While our primary emphasis here is on synthetic scenarios, this approach is adaptable to real-world cases, where arrival times are obtained from seismogram recordings. In this case, to generate new travel-time and velocity datasets for training a modified ANN, the corresponding models would need to incorporate realistic 3D topography and a source-receiver configuration based on actual seismological field studies.

## 3    Results

We consider two different source-receiver configurations. Both are characterized by sparse ray coverage, where the linear inversion fails to yield satisfactory results. By optimizing the ANN, we aim to achieve more accurate inversion models. In the first case, sources are placed on one side, and a dense array of receivers is placed on the opposite side, simulating an active source seismic experiment. In the second case, we consider earthquakes located within the volcanic edifice. In both cases, we invert the calculated travel times of the input model to reconstruct the velocity model using both damped linear least-squares inversion and the ANN.

We further consider two types of anomalies with respect to a uniform background velocity model of 1.5 km/s: i) a single anomaly of non-regular shape comprised of 7 blocks with a velocity of 1 km/s (as shown in Figure 3a) and ii) three square-shaped anomalies, where each square is composed of 4 blocks with a velocity value of 1 km/s. These multiple anomalies are distributed within the edifice as illustrated in Figure 7a.

### 3.1 ANN optimization

We commence by simulating an active source seismic experiment, utilizing a model that includes a single velocity anomaly. This setup is used to optimize our ANN approach. Three sources are placed along one flank of the edifice while the 81 receivers are placed on the opposite side (Figure 3a). A recent example for a similar real setup to study the Erebus volcano is described in Zandomeneghi et al. (2013) with 12 chemical shots and 91 seismic receivers. Their study highlighted resolution constraints related to the imaging of conduit structures beneath the inner crater using conventional tomography techniques.

The damped linear inversion velocity result is depicted in Figure 3b. Note that only the blocks that were involved in the inversion process are shown (blocks that are not traversed by any of the rays are omitted). The travel times obtained through the forward process using the velocity model of the linear inversion perfectly match the input model travel times, as expected from equation 3.

However, it becomes evident that the linear inversion failed to retrieve the velocity model. Extensive smearing along the raypaths obscures the central low-velocity anomaly, making it indistinguishable from surrounding spurious low and high velocities. Moreover, the velocities in the model resulting from the inversion are predominantly at the extremes of 1 and 2 km/s. This issue becomes even more pronounced when a gradient background velocity is employed (see Figures S4 and S5).

#### 3.1.1 Effects of optimizers on ANN training

During the trial and error process to determine the hyperparameters that are best suited for the inversion, we observed that employing different optimizers could result in notable variations in the inversion results (Figures 4 and 5). These optimizers fall into two categories: first-order (e.g., Rprop, SGD, Adam, RMSProp) and second-order optimizers, such as L-BFGS (see also Kashyap, 2022 for a general review). The order of an optimizer is determined by its approach to optimize the objective function (i.e., loss function in machine learning context). Two main approaches to optimize loss function are gradient descent and Newton methods (Boyd and Vandenberghe, 2004). The former, to update the function's parameters, takes the steps towards the optimum point in the negative direction of the gradient, and in the latter these steps are normalized by second-order derivatives of the function relative to its variables. When the second-order derivatives are used as preconditioning in optimization problems, particularly in cases where the function is convex, faster and more precise convergence to the optimum point can be achieved. The optimization process can be expressed as

$$m^{n+1} = m^n - \eta H^{-1} \nabla(L(m)) \quad (5)$$

where $m$ is a vector of model parameters, $\eta$ is the learning rate, $H$ is the Hessian matrix, which holds the second-order derivatives of the loss function relative to the model parameters, and $\nabla(L(m))$ is the gradient of the loss function. In view of the additional computational burden imposed by second-order derivatives, in huge neural networks it may not be practical to use these optimizers. However, within the scope of our problem, maintaining a simple neural network structure facilitates the use of the L-BFGS optimizer. The L-BFGS is a quasi-Newton optimizer that employs an approximation of the Hessian matrix (e.g., Liu and Nocedal, 1989).

We compared the performance of our proposed ANN using various first-order optimizers against the L-BFGS optimizer, as shown in Figure 4. The L-BFGS optimizer managed to reduce the loss function to a target value within just a few epochs, a feat that first-order optimizers typically took about 200 epochs to achieve. By the 1500th epoch, the L-BFGS optimizer recorded a training loss value of 0.0098. In contrast, the best result from first-order optimizers was 0.0158, achieved using the Adam optimizer after 10,000 epochs (see Figure 4a). It is noteworthy that reaching 1500 epochs with the second-order L-BFGS and 10,000 epochs with the first-order optimizers required the same computational time. Figure 4 (c and d) displays the loss function values of the validation dataset across epochs. These graphs indicate that the loss function of the validation dataset follows the pattern seen in the training dataset over the epochs. Importantly, the loss function values for the validation dataset are slightly elevated compared to the training dataset, which suggests that overfitting is unlikely. The velocity models (and corresponding travel times) derived from the ANN inversion using the L-BFGS and the different first-order optimizers are presented in Figure 5. Only the Adam optimizer produces results that may be considered similar to the input velocity model and to the results obtained using L-BFGS.

Given our simplifying assumptions, we do not anticipate that more complex architectures, such as Recurrent Neural Networks (RNNs) or Residual Networks, would be suitable for our problem Nevertheless, we investigated the effect of architecture design on our problem by adjusting the number of hidden layers and neurons. Figures S2 and S3 demonstrate that adding hidden layers, convolutional layers, or increasing the number of neurons can delay the convergence to the optimal point of the loss function. Conversely, attempts to identify velocity anomalies using conventional machine learning approaches, such as Random Forests and Support Vector Machines (SVM), were unsuccessful (Figure S6), possibly due to the high dimensionality of the problem. Notably, these methods also require more training time compared to the ANN.

### 3.1.2. Inversion accuracy evaluation

As described in section 2.3, we employ a test dataset consisting of 1500 samples, comprising travel times and their corresponding velocity models. These input-output pairs have not been exposed to the ANN during the training phase. To quantitatively assess the performance of the ANN and linear inversion methods in retrieving these previously unseen models, we use two different error evaluation methods: the Root Mean Squared Error (RMSE) and the Structural Similarity Index Metric (SSIM, Wang et al., 2009). For both methods, we utilize open-source Python modules from the scikit-image library. When calculating RMSE, the input and inversion velocity models are treated as arrays. In the case of SSIM, these models are regarded as images. According to Wang et al. (2009), SSIM incorporates image features such as structure, contrast, and luminance. Figure 6 displays histograms that depict the error estimations for ANN inversion (a and c) and linear inversion (b and d). Note that in the SSIM method, similarity between two images is represented by values ranging from 0 to 1, with 1 indicating complete similarity.

These histograms quantitatively demonstrate that the ANN outperforms linear inversion results, showing the velocity models retrieved by the ANN are much more similar to the corresponding input models. The mean RMSE for the entire test dataset is approximately 0.1 s/km for the ANN inversion (Figure 6a), whereas it is about 0.7 s/km for the linear inversion (Figure 6b). Notably, the latter value is even greater than the difference between the anomalous medium and the background velocity, which is around 30%. Additionally, the SSIM histograms (Figure 6c and 6d) indicate higher values for the ANN inversion, suggesting a closer resemblance between the input model and the ANN inversion model. The SSIM values for the ANN inversion are around 0.98 (Figure 6c), compared to about 0.85 for the linear inversion (Figure 6d). Table 1 provides the error estimation for the input velocity model shown in Figure 3a, along with additional source-receiver configurations, which will be discussed in the following section. Furthermore, Table S1 presents the same error evaluations for ANN inversion using both the L-BFGS and four different first-order optimizers, as depicted in Figure 5.

### 3.2 Alternative velocity model and source-receiver configuration

After optimizing the ANN, we evaluate its performance using an alternative input velocity model featuring multiple anomalies while maintaining the same source-receiver setup (i.e., the active-source simulation). Figure 7 illustrates the related input velocity model, the linear inversion result, the ANN inversion result, and the corresponding travel times obtained via forward modeling using both the input and ANN inversion velocity models. In this scenario with multiple anomalies, our results demonstrate that the ANN inversion surpasses the linear inversion in terms of accuracy. The velocity anomalies can be readily distinguished in the ANN inversion, while in the linear inversion, they are difficult to discern, with anomalies tending to spread along the raypaths, particularly near the sources. However, it is worth noting that the ANN inversion also presents some elevated velocities around the anomalies. These higher velocities are mostly within 5-10% and can reach up to 20% in certain blocks relative to the background velocity. These artifacts become more conspicuous in the presence of multiple velocity anomalies.

Imaging deeper parts of the edifice's subsurface and conducting (near) real-time monitoring of volcanoes to assess related hazards pose challenges for active experiments (e.g., Kiser et al., 2019; Shalev et al., 2010). In such cases, seismograms from earthquakes occurring within the edifice could be utilized to illuminate the internal structures. Such earthquakes can be caused, for example, by magnetic fracturing (Rümpker et al., 2022). We present results for earthquakes located within the edifice, addressing both single and multiple velocity anomalies in Figures 8 and 9. In this scenario, three sources were placed inside the deeper parts of the edifice and 71 receivers were positioned on both sides. Note that, in this case, the ray coverage is sparser compared to the previously simulated active-source experiment. However, for a better comparison, we kept the input velocity models unchanged see Figures 8a, 9a). The inverted velocity models from linear inversion and ANN methods are illustrated in Figures 8b, c and 9b, c.

The linear inversion models suffer from a high degree of smearing along the raypaths. Additionally, since the anomalies are close to the sources, their detection becomes impossible. In contrast, Figures 8c and 9c demonstrate the remarkable capacity of the ANN to resolve the velocity structures present in the input model. Nonetheless, the ANN results are influenced by the less ray coverage, which leads to slight smearing along the raypaths and the higher-velocity artifacts (maximum ~5%) around the anomalies. The corresponding travel times resulting from the ANN inversion models are presented in Figures 8d and 9d, in comparison to those obtained from the input model. As evident from these figures, the travel times from the ANN inversion models show good agreement with those resulting from the input velocity models.

## 4 Discussion and Conclusions

In this study, we developed an effective Artificial Neural Network (ANN) that utilizes a second-order derivative optimizer (L-BFGS) to resolve anomalous velocity structures within a volcanic edifice based on provided travel times. We evaluated the performance of this network by comparing its ability to resolve the velocity models against that of a conventional linear damped least-squares inversion method. The ANN was trained on synthetic pairs of travel times and randomly generated velocity models. After its training phase, the ANN was able to successfully predict the velocity model for a new set of travel times. We demonstrated the superior performance of the ANN over the linear inversion method through the simulation of two scenarios involving (i) an active source experiment and (ii) earthquakes located within the edifice. With the sparse ray coverage addressed in our examples, conventional damped least squares inversion led to unclear velocity models, often yielding highly smeared solutions and failing to accurately resolve inhomogeneous subsurface structures.

We employed the L-BFGS optimizer in our study to minimize the loss function, particularly leveraging its capability to utilize second-order derivatives for updating the weights of the neural network. Our comparative analysis of the efficiency of the L-BFGS optimizer in minimizing the loss function, in comparison to other commonly used first-order optimizers such as Adam and RMSProp, highlighted its strengths in terms of convergence speed and accuracy when reaching the optimum point. However, it is important to note that the use of second-order derivatives, while advantageous, is only practically feasible in relatively small neural networks due to the higher computational burden.

One significant advantage of the ANN over traditional tomography methods is its independence from an initial velocity model. While the training of the ANN with thousands of velocity models might require more computation time than traditional tomography methods, upon completion of the training, the ANN can rapidly perform inversions for any new set of travel times. In this study, we utilized mesh-dependent velocity models for training the ANN to allow for comparison between the ANN and linear inversion results. However, it is also feasible to use mesh-free velocity models and take advantage of high-resolution images of velocity structures to train the neural network. We did not conduct tests for different resolutions (i.e., block size) of the velocity structure, which is a factor that needs to be taken into account in future studies and for real-case scenarios.

Sparse-ray tomographic inversions, as facilitated by ANN implementations, may offer new possibilities for volcano monitoring. Unlike traditional linear and non-linear tomographic methods that require a significantly larger number of sources and receivers (e.g., Lees, 2007), our findings suggest that the ANN approach can effectively resolve anomalies and their potential changes even with sparse ray coverage, which can be maintained over extended periods.

In comparison to previous studies (e.g., Araya-Polo et al., 2018; O'Brien et al., 2023), our approach emphasizes the advantages of using second-order derivatives especially in situations with sparse ray coverage. It is important to emphasize that in this study, we utilize random velocity models as the output for the neural network during the training phase, which differs significantly from the final example velocity models applied to the ANN. This sets our approach apart from previous studies where the training velocity models were smoothed versions of the final testing examples (e.g., O'Brien et al., 2023).

Regarding the simplified assumptions we have adopted in the forward problem, it is essential to consider their potential implications for the final inversion results. In real-world situations, waves traverse through heterogeneous media and encounter phenomena such as reflection, refraction, and diffraction. Techniques for determining travel times in such scenarios are available and can be implemented for modeling more complex velocity structures. Another simplification relates to assuming that the source locations within the edifice are known, a priori. A trade-off between the velocity model and the earthquake location that influences the travel times can be expected. This may require a relaxation of the earthquake location in a joint inversion approach, to invert for both the velocity models and the earthquake location. It is also worth noting the potential for data contamination by various noise sources. In these circumstances, additional layers of the ANN, possibly convolutional, might be necessary to discriminate relevant travel-time patterns effectively from the noisy data. Lastly, while our calculations are presently in 2D, they can be extended to 3D in subsequent studies. This expansion, however, will likely require significantly more training examples for neural networks to identify features within the model space effectively.

The inversion models resulting from the linear approach, in both scenarios, are highly contaminated by extreme smearing along the raypaths. This contrasts with the ANN approach, which has well resolved the anomalies in both scenarios. However, there are several strategies that could potentially improve the performance of the ANN. A promising direction for future research could be to explore physics-informed Neural Networks (PINN) methods. These methods would combine the robustness and simplicity of linear methods with the power and flexibility of neural networks. This could potentially result in more accurate and stable inversion results, further advancing the field of tomographic inversion. Moreover, incorporating more realistic assumptions, such as employing the wave equation for forward modelling in both 2D and 3D, or utilizing the full-waveform instead of travel times for training the ANN, could enhance the capability of the ANN to handle real datasets in future research.

**Permission to reuse and copyright**

Permission must be obtained for use of copyrighted material from other sources (including the web).

**Conflict of Interest**

*The authors declare that the research was conducted in the absence of any commercial or financial relationships that could be construed as a potential conflict of interest.*

**Author Contribution**

GR, AK and FL conceptualized the study. AK and GR performed the calculations and wrote the first draft of the manuscript. JF and NS contributed to designing the ANN. All authors contributed to manuscript revision, read, and approved the submitted version.

**Funding**

NS is supported by the "KI-Nachwuchswissenschaftlerinnen" - grant SAI 01IS20059 by the Bundesministerium für Bildung und Forschung - BMBF.

**Acknowledgments**

AK acknowledges support by a Goethe-University scholarship provided through the Department of Geosciences/Geography. NS expresses gratitude for the support received from the Bundesministerium für Bildung und Forschung - BMBF.

**Data Availability Statement**

All (synthetic) data utilized in this study were generated using the codes described in the text. The corresponding datasets can be re-computed following the methodology outlined in the manuscript.

**Figures**

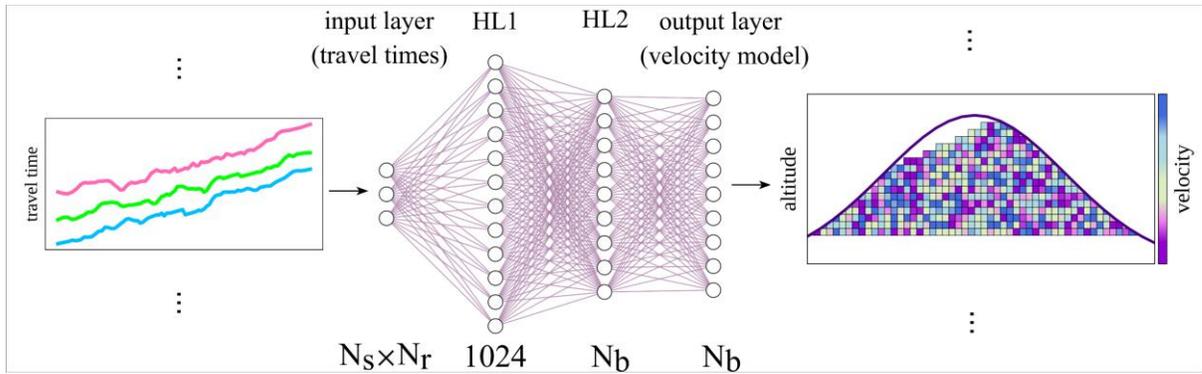

Figure 1. Schematic architecture of the proposed ANN. We consider cases of fewer neurons in the input layer in comparison to the output layer which corresponds to an ill-posed inversion problem, where the number of rays is less than the number of model parameters ($N_s*N_r < N_b$). This Figure depicts a typical pair of travel times and corresponding velocity model used in training the ANN. Ns denotes the number of sources, Nr the number of receivers and Nb the number of blocks or model parameters. HL denotes a hidden layer.

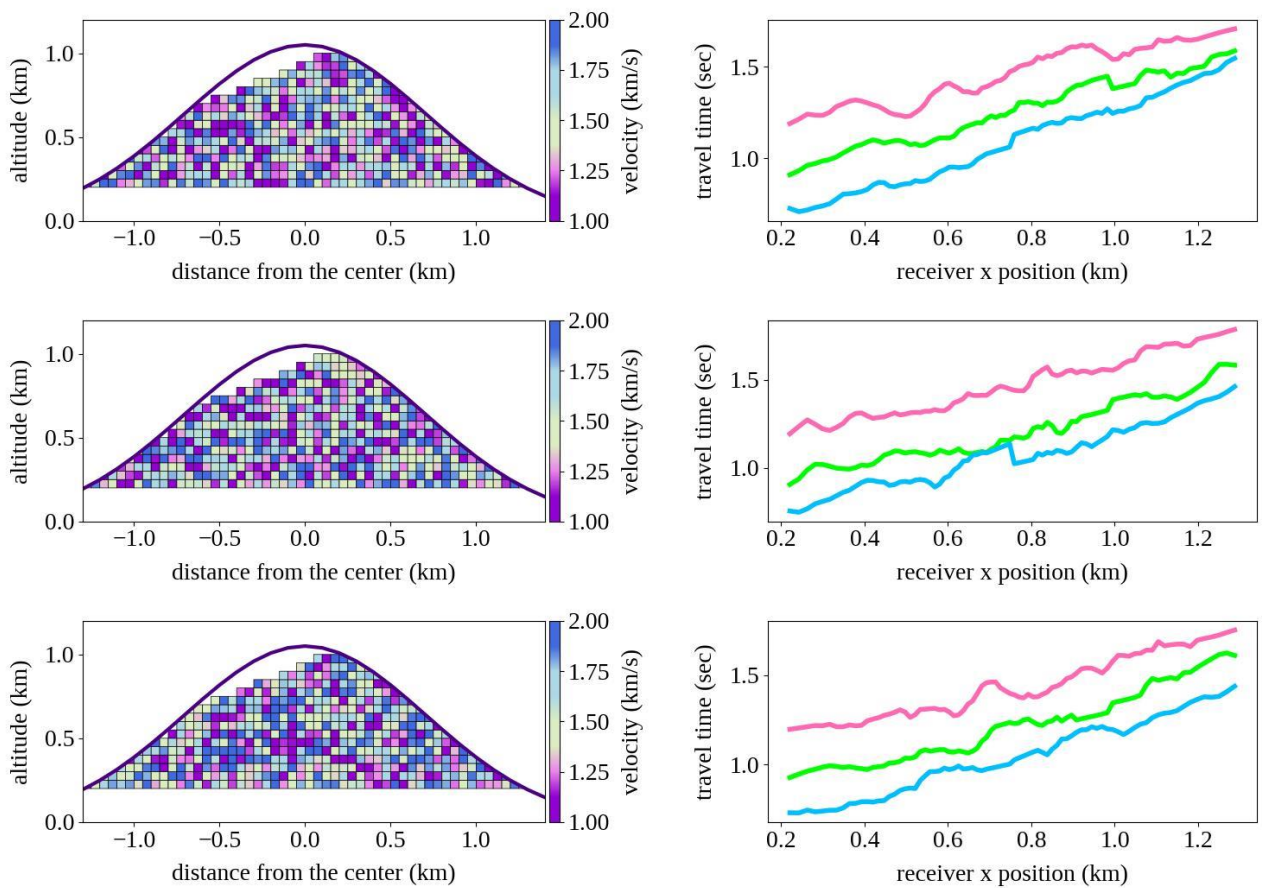

Figure 2. Typical examples of velocity models (left panel) and corresponding travel times (right panel) used for training the neural network. The travel times (training inputs) are calculated from the corresponding velocity models (training outputs) via forward modeling (refer to section 2.1). For a given model, velocity values in each block were chosen randomly between 1 and 2 km/s.

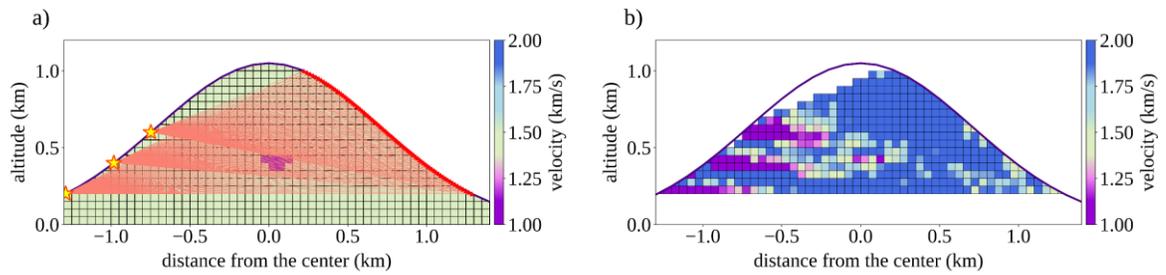

Figure 3. Illustration of the active source experiment setup and corresponding inversion results. a) Three sources are located on the left-hand side and 81 receivers on the right-hand side of the edifice. The velocity model features a single velocity anomaly of 1 km/s in a background velocity model of 1.5 km/s, b) Result of the damped least-squares inversion.

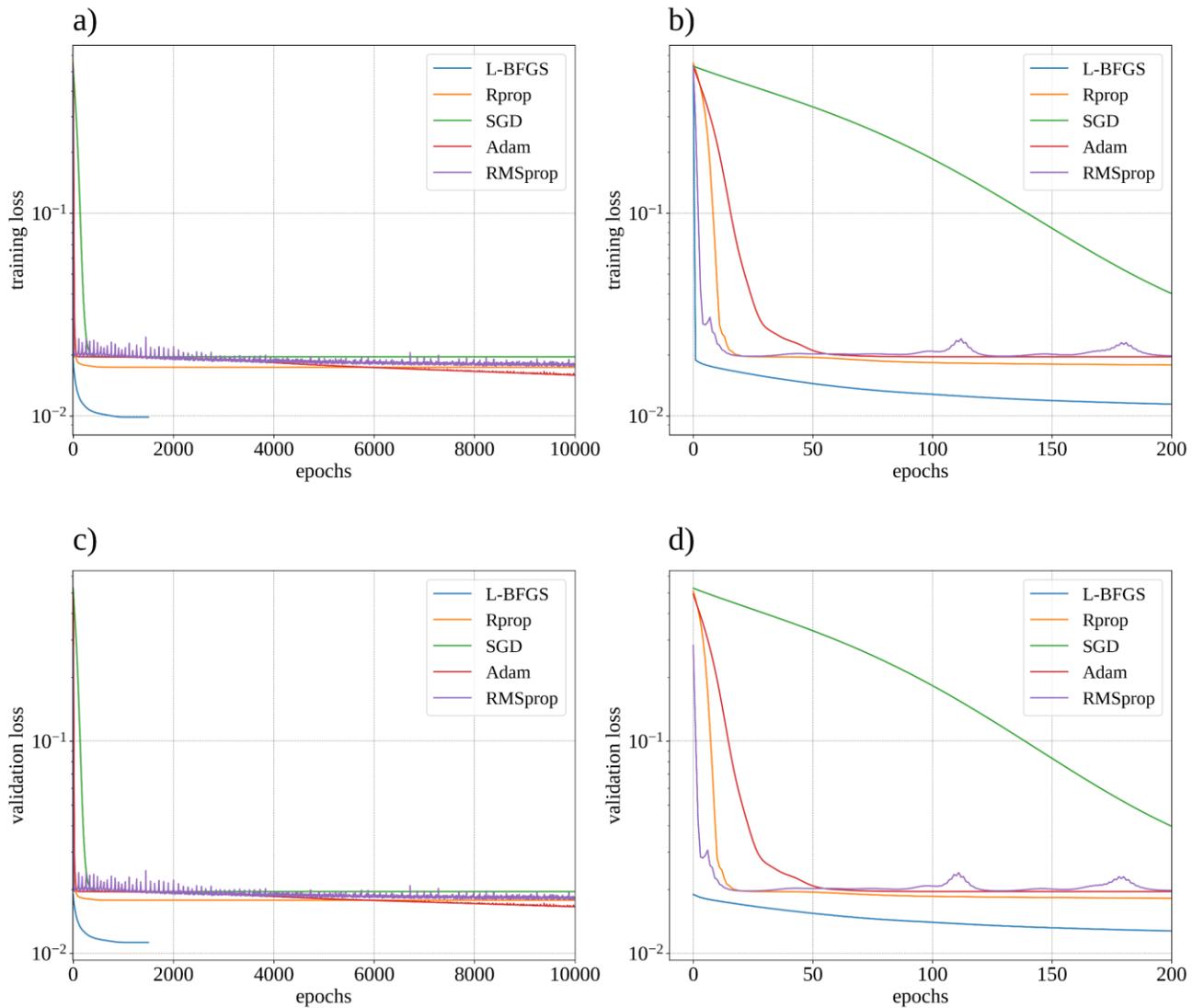

Figure 4. Comparison of training loss using the L-BFGS as optimizer versus different first-order optimizers. a) The training loss is displayed for all epochs (10000 epochs for the first-order optimizers and 1500 epochs for the L-BFGS). b) Zoomed-in view of the initial 200 epochs from the graph depicted in a. c) The validation loss is depicted for all epochs (10000 epochs for the first-order optimizers and 1500 epochs for the L-BFGS). d) Zoomed-in view of the initial 200 epochs from the graph depicted in c.

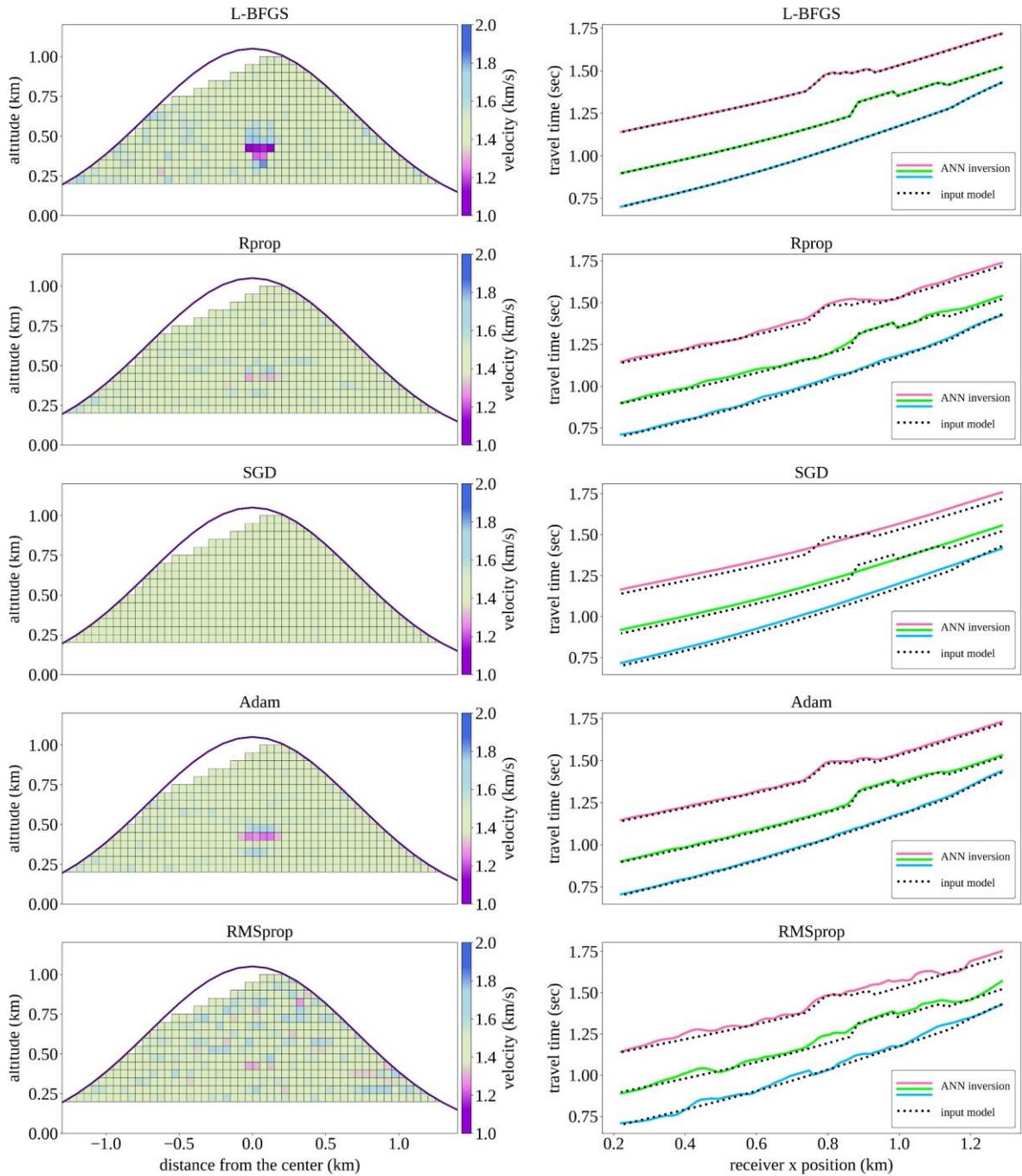

Figure 5. Left panel: velocity models resulting from ANN inversions using the second-order L-BFGS and four different first-order optimizers. Right panel: the corresponding travel times for the models shown on the left in comparison with the travel times of the input model.

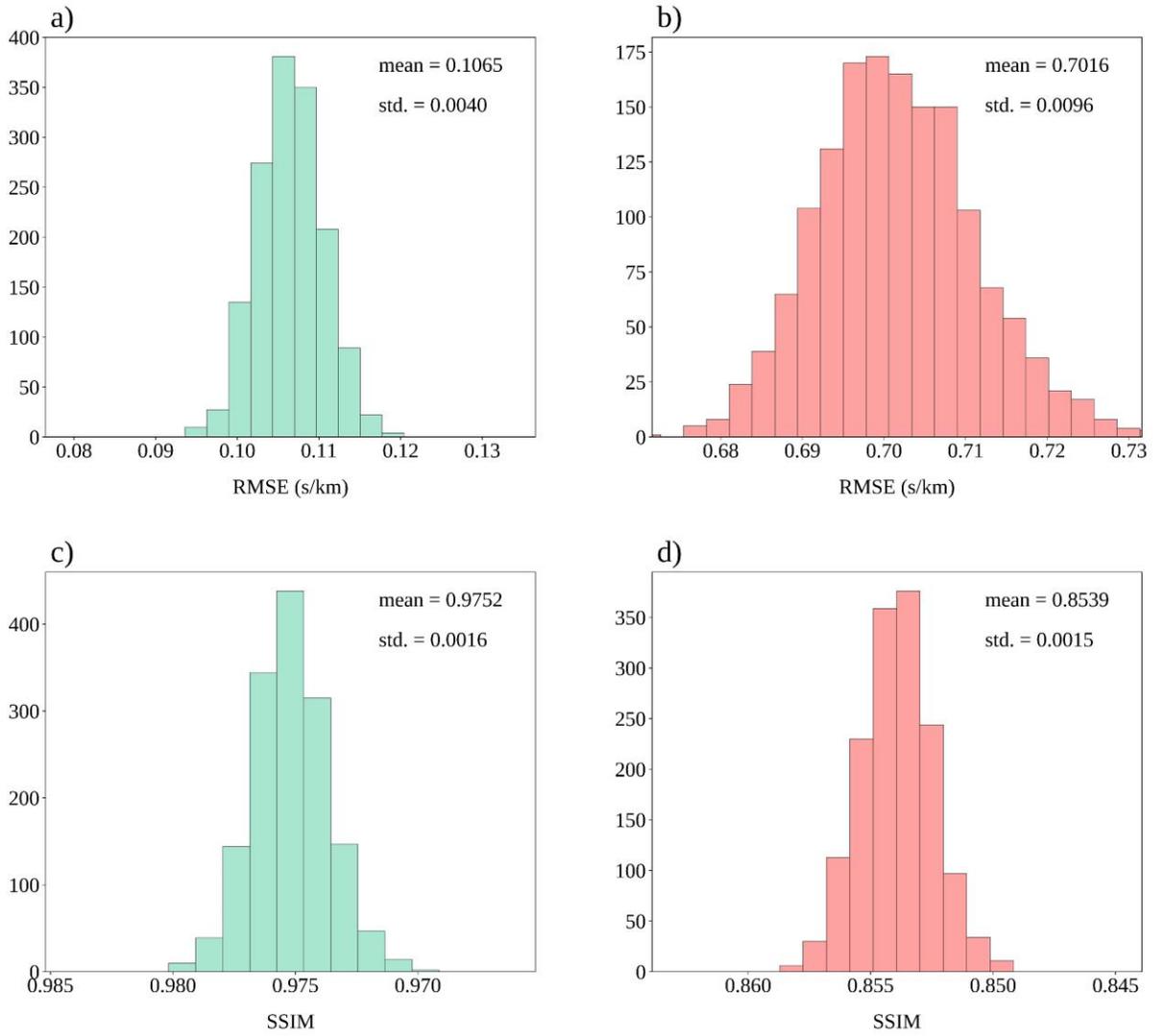

Figure 6. Error evaluation of the ANN (a and c, green histograms) and linear inversion (b and d, red histograms) methods performance on the test dataset. RMSE, stands for root mean squared error and SSIM stands for structural similarity index metric.

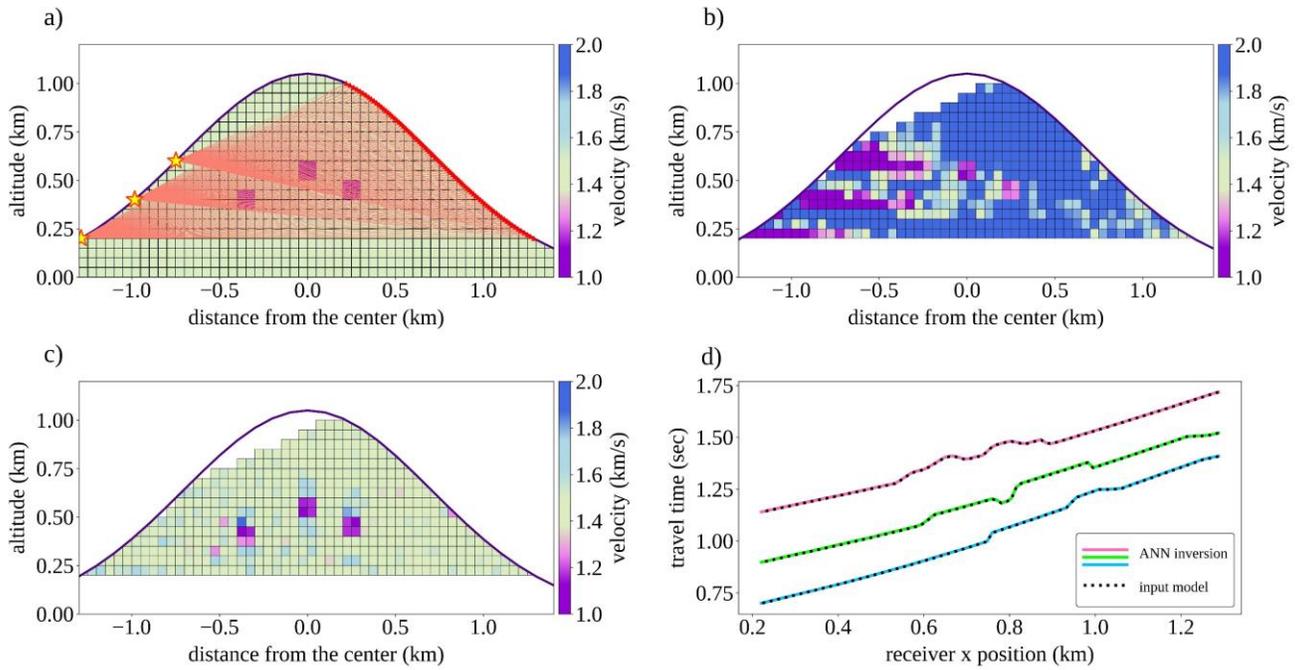

Figure 7. Illustration of the active source experiment simulation setup (with multiple anomalies) and inversion results. a) setup with 3 sources on the left-hand side and 81 receivers on the right-hand side of the edifice. The input model features multiple velocity anomalies of 1 km/s in a background velocity model of 1.5 km/s, b) linear damped least-squares inversion velocity model, c) ANN inversion velocity model, d) comparison of travel times resulting from the input model (black dotted line) and ANN inversion (colored solid lines) for each source.

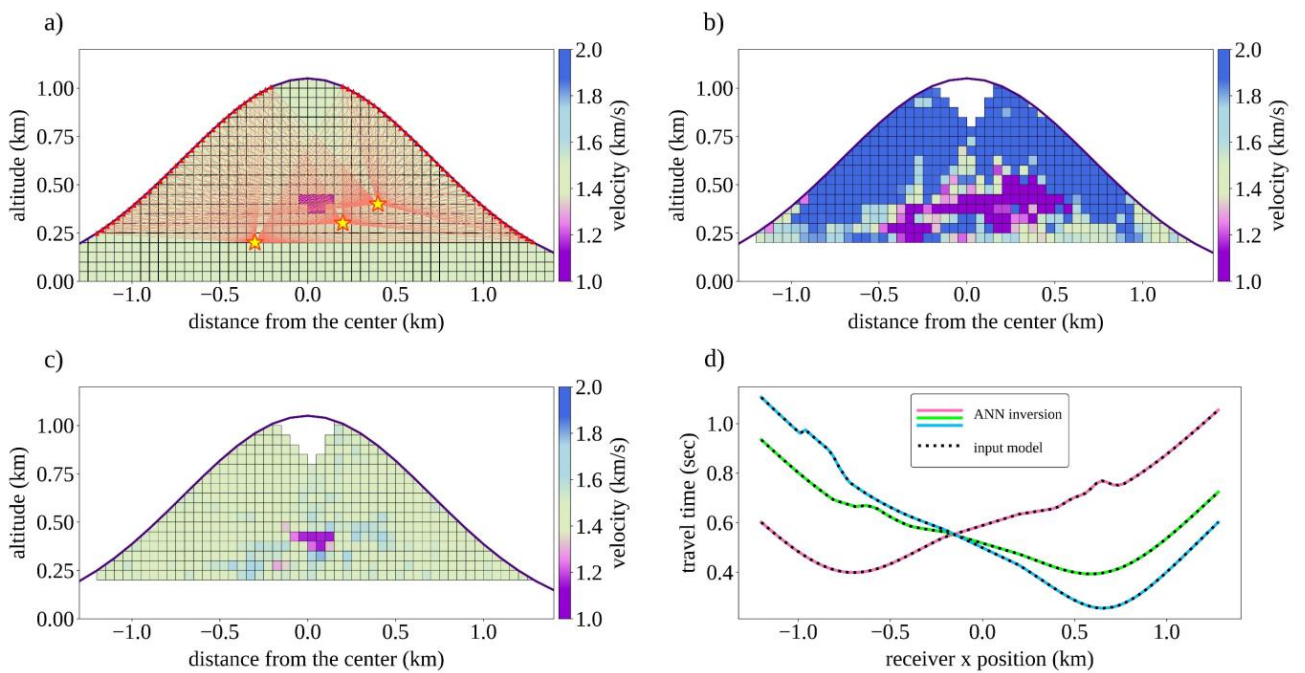

Figure 8. Illustration of the edificial earthquake simulation setup and inversion results. a) The simulation setup including 3 sources located within the edifice and 71 receivers on both sides of the edifice. The input velocity model exhibits a single velocity anomaly of 1 km/s in a background velocity model of 1.5 km/s, b) linear damped least-squares inversion velocity model, c) ANN inversion velocity model, d) comparison of travel times resulting from the input model (black dotted line) and ANN inversion (colored solid lines) for each source.

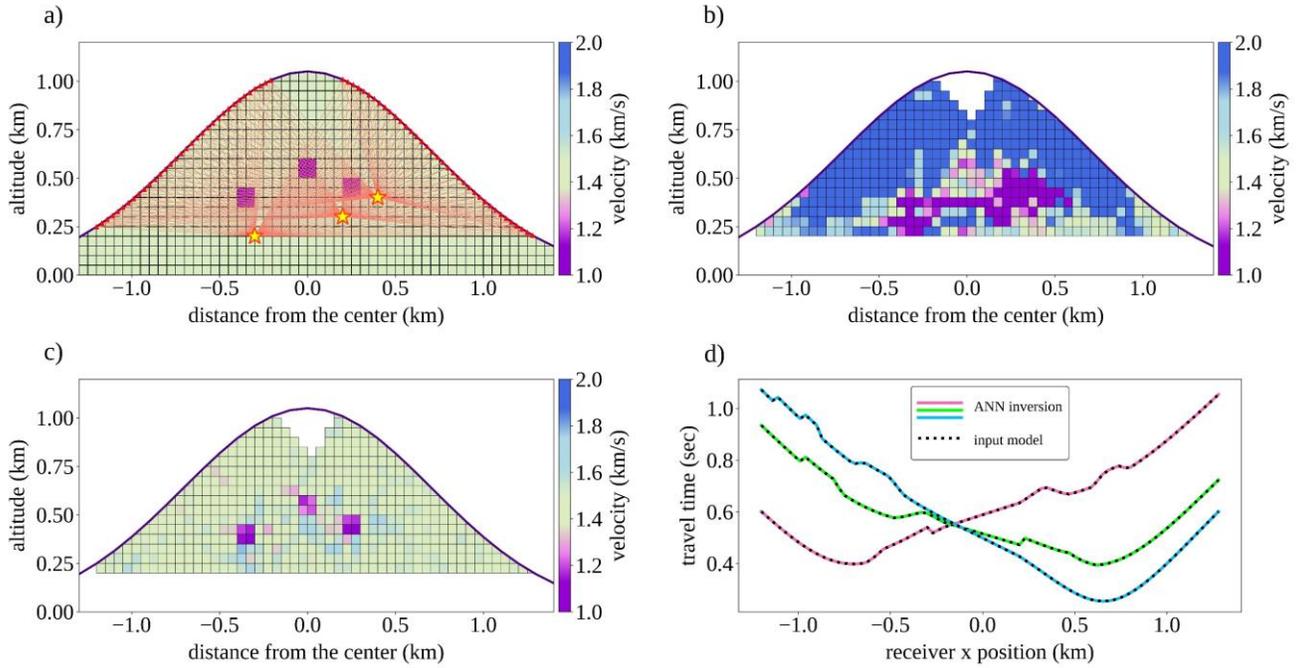

Figure 9. Same as Figure 8, with three distinct velocity anomalies as shown in a (input model).

| source-receiver conf. | active source simulation | | | | edificial earthquake simulation | | | |
|---|---|---|---|---|---|---|---|---|
| type of anomaly | single | | multiple | | single | | multiple | |
| inversion method | ANN | linear | ANN | linear | ANN | linear | ANN | linear |
| RMSE | 0.028 | 0.677 | 0.038 | 0.305 | 0.029 | 0.325 | 0.038 | 0.325 |
| SSIM | 0.998 | 0.850 | 0.997 | 0.849 | 0.998 | 0.851 | 0.997 | 0.851 |

Table 1. Error evaluation of the ANN and linear inversion methods, for all the source-receiver configurations, and different type of anomalous mediums, used in this study. Active source simulations for single and multiple anomalous media are depicted in Figure 3a and Figure 7a, respectively. Figure 8a and Figure 9a illustrate the edificial earthquake simulation for single and multiple anomalous media, respectively.

# When linear inversion fails: neural-network optimization for sparse-ray travel-time tomography of a volcanic edifice

**Abolfazl Komeazi[1*], Georg Rümpker[1,2], Johannes Faber[2], Fabian Limberger[1], Nishtha Srivastava[1,2]**

[1]Institute of Geosciences, Goethe University Frankfurt, Frankfurt, Germany
[2]Frankfurt Institute for Advanced Studies, Frankfurt, Germany

**\* Correspondence:**
Abolfazl Komeazi
komeazi@geophysik.uni-frankfurt.de

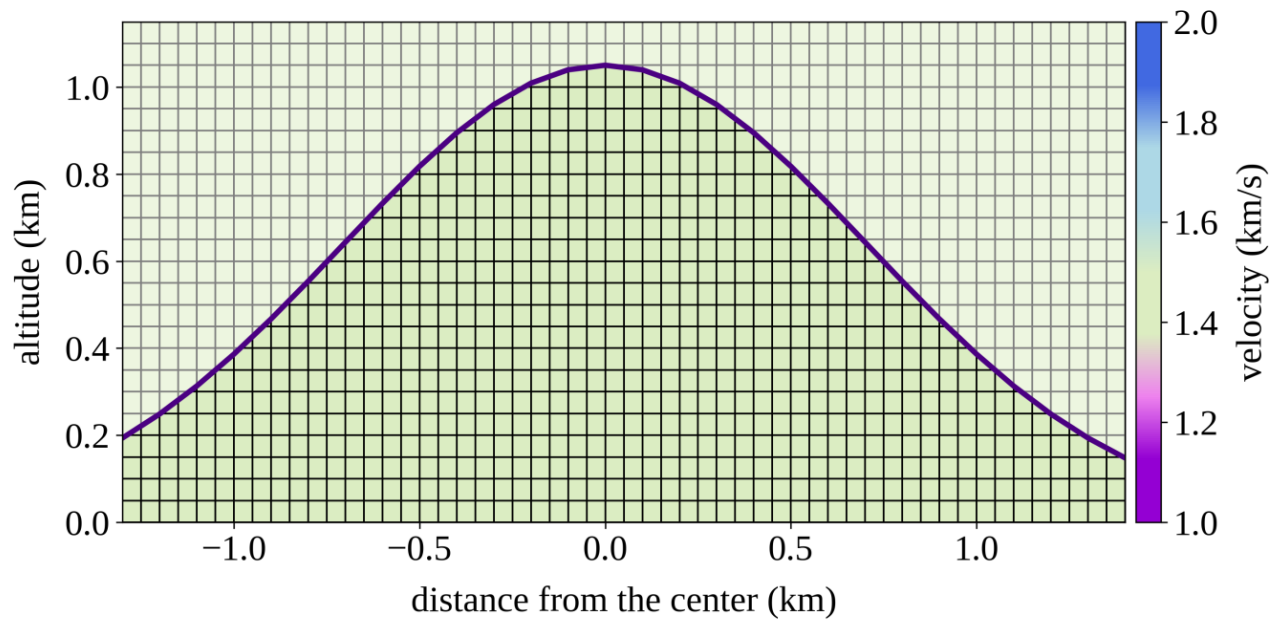

Figure S1. Edifice model setup. A gridded rectangle with dimensions of 2.7 km in length and 1.15 km in height, the grid points are spaced at intervals of 0.05 km in both the x and z directions. The shape of the edifice corresponds to a Gaussian function.

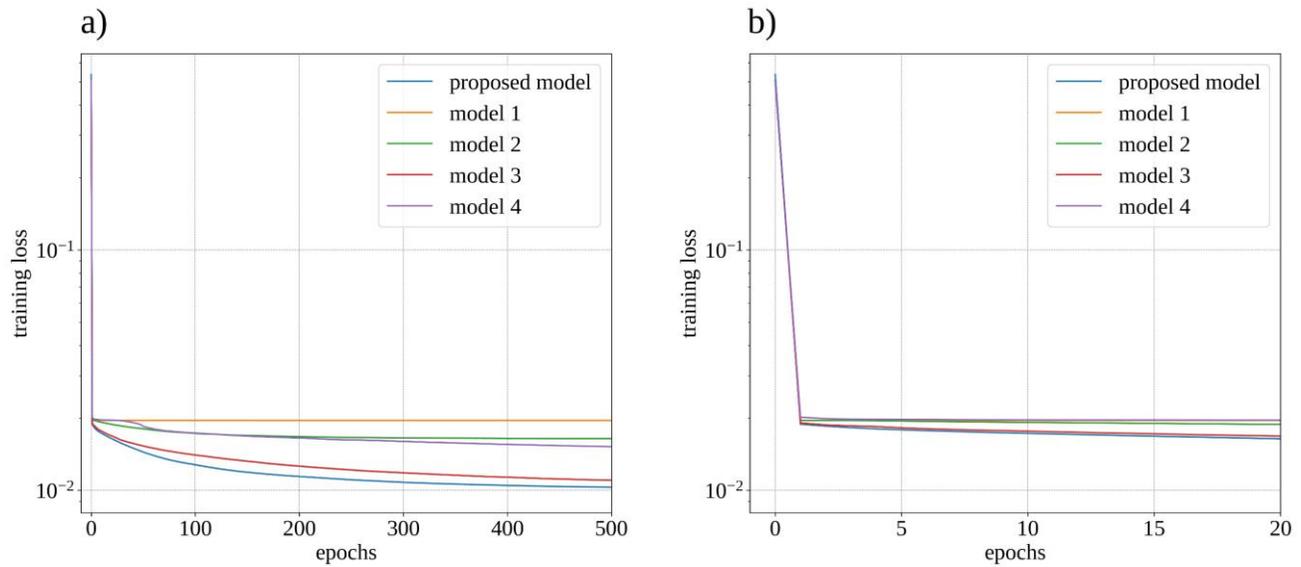

Figure S2. a) Illustration of training loss across 500 epochs for various architectures employed to resolve the velocity model in a scenario with 3 side sources and 81 receivers. b) Zoomed-in view of the initial 20 epochs from graph a. Models 1 and 2 represent architectures with more hidden layers with increasing and decreasing the dimension of the input layer. Model 3 is the same as our proposed ANN with an additional hidden layer (1024 neurons), and model 4 includes convolutional layers.

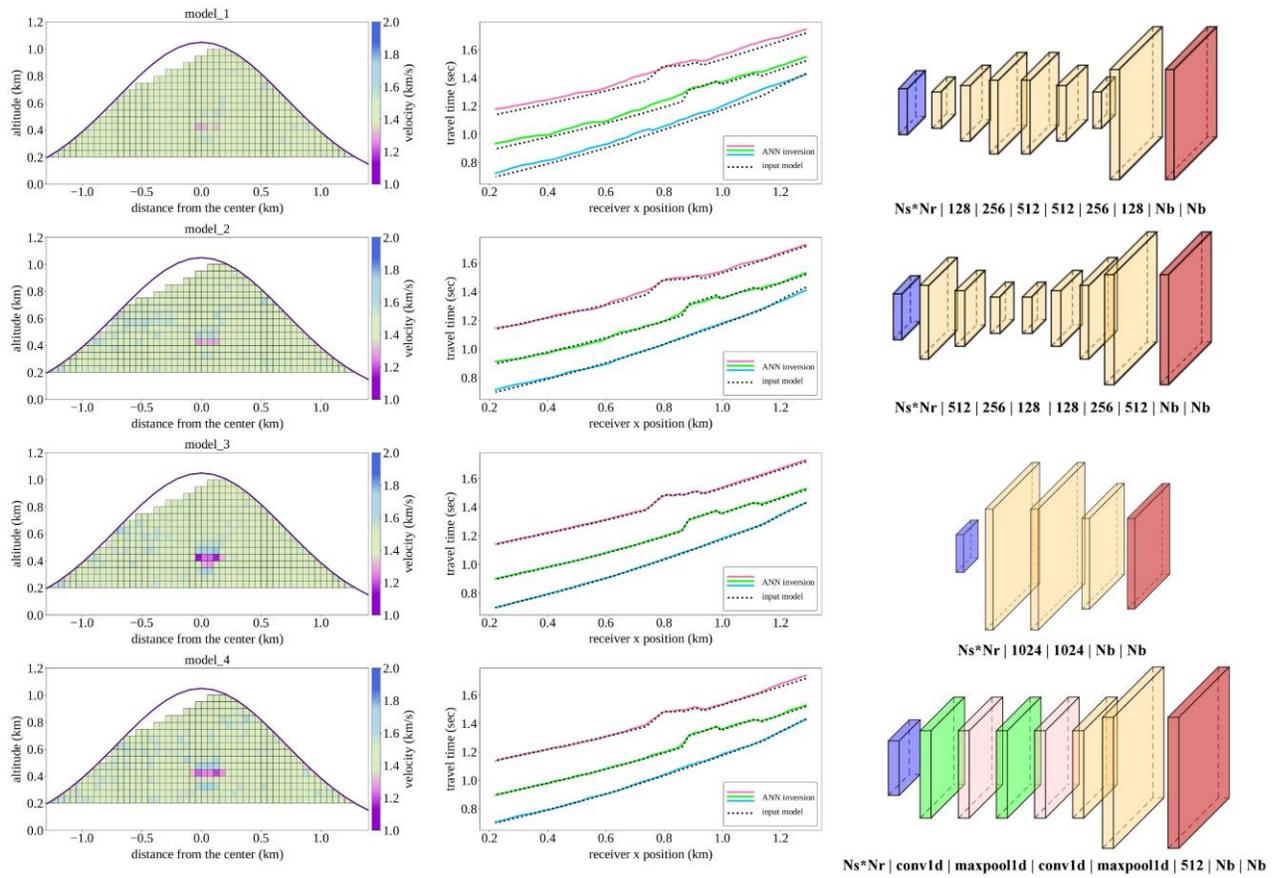

Figure S3. The left column illustrates the velocity models resulting from the ANN inversion with the architecture depicted in the right column. The corresponding travel times are shown in the middle column. Models 1 and 2 represent architectures with more hidden layers with increasing and decreasing the dimension of the input layer. Model 3 is the same as our proposed ANN with an additional hidden layer (1024 neurons), and model 4 includes convolutional layers. Ns denotes the number of sources, Nr the number of receivers, and Nb number of blocks or model parameters. Numbers in the right column represent the number of neurons in each hidden layer.

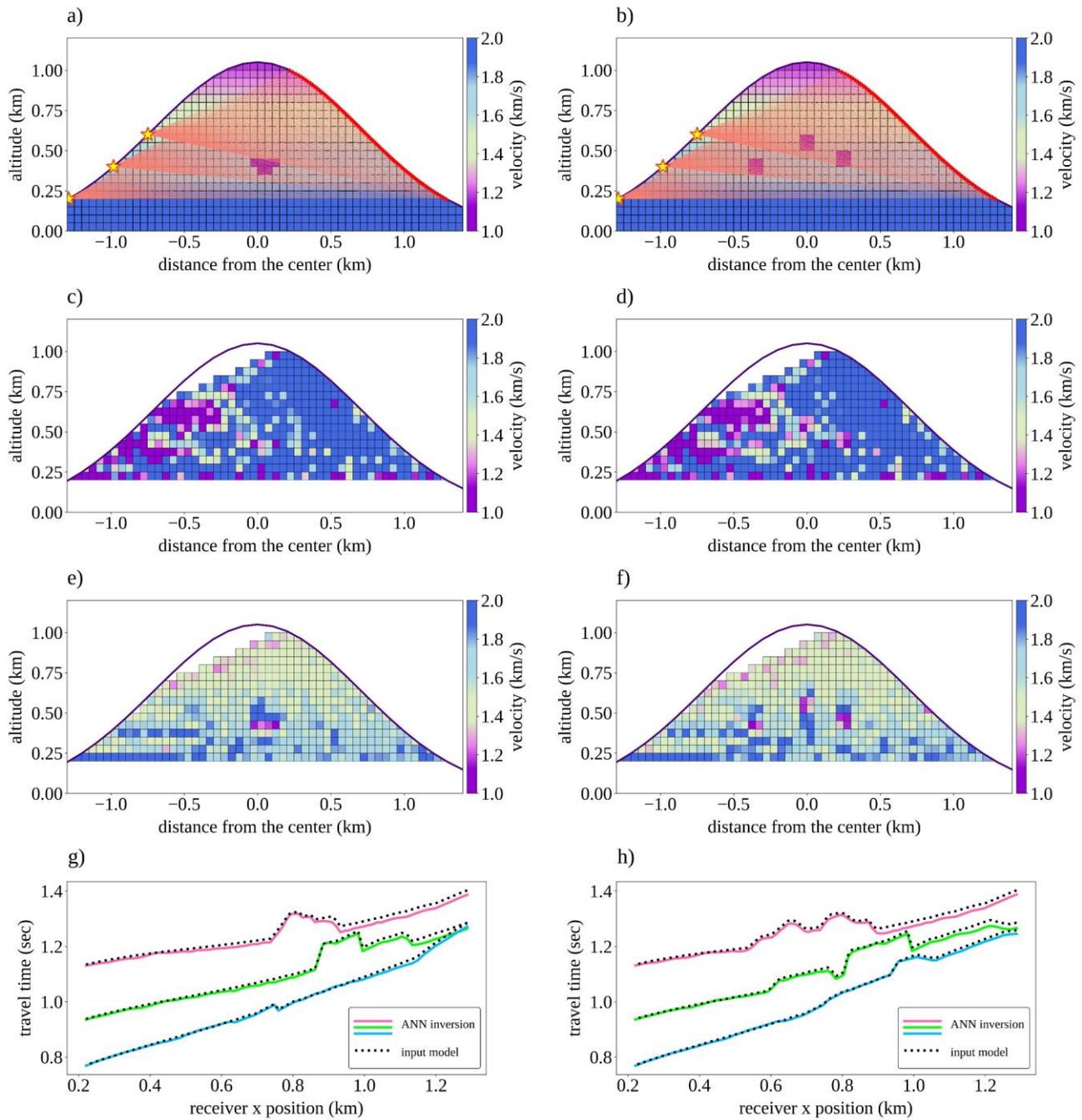

Figure S4. Illustration of the active source experiment simulation setup (with 3 source and 81 receivers) and inversion results by having a gradient background velocity model. a) The input velocity model features a single velocity anomaly of 1 km/s in a gradient background velocity model changing from 1.1 to 2 km/s, b) the input velocity model features three distinct velocity anomalies of 1 km/s in a gradient background velocity model changing from 1.1 to 2 km/s, c,d) linear damped least-squares inversion velocity model, e,f) ANN inversion velocity model, g,h) comparison of travel times resulting from the input model (black dotted line) and ANN inversion (colored solid lines) for each source.

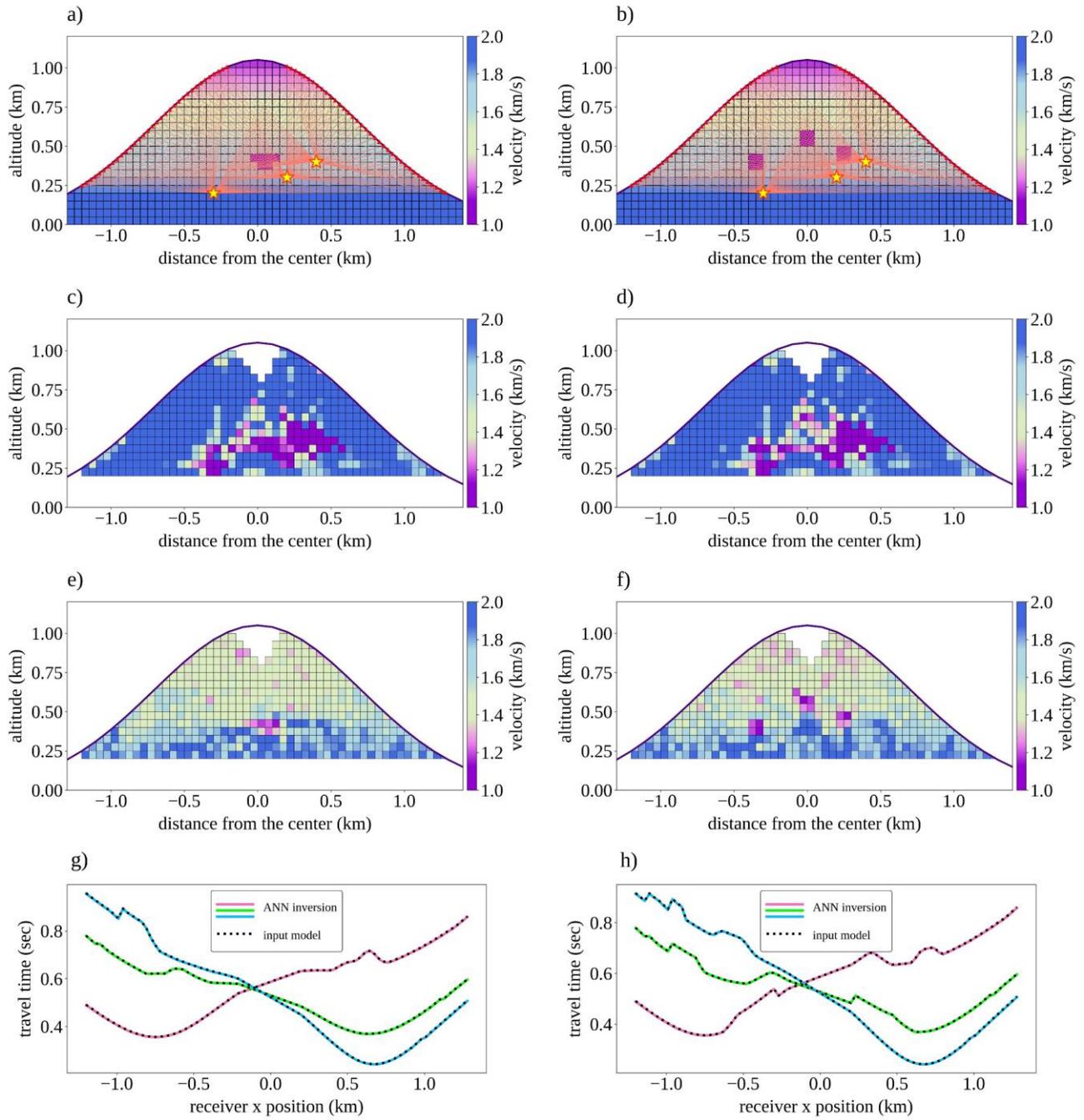

Figure S5. Same as Figure S4 for the edificial earthquake simulations, with 3 sources and 71 receivers.

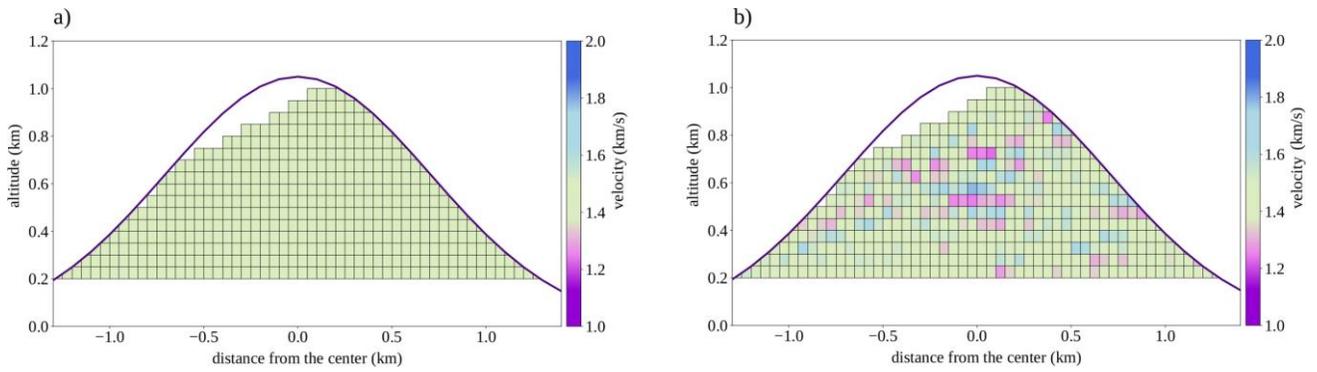

Figure S6. Inversion results using a) random forests and b) Support Vector Machine (SVM) approaches. The input model (true model) is presented in Figure 3a.

| optimizer | RMSE | SSIM |
|---|---|---|
| L-BFGS | 0.028 | 0.998 |
| Rprop | 0.036 | 0.987 |
| SGD | 0.038 | 0.987 |
| Adam | 0.032 | 0.991 |
| RMSprop | 0.045 | 0.987 |

Table S1. Error evaluation of the ANN (marked with green color) using the second-order L-BFGS and four different first-order optimizers for the configuration shown in Figure 3a. The corresponding inversion velocity models are illustrated in Figure 5.